# Fixed-point iterative algorithm for SVI model

Shuzhen Yang* Wenqing Zhang†

**Abstract**

The stochastic volatility inspired (SVI) model is widely used to fit the implied variance smile. Presently, most optimizer algorithms for the SVI model have a strong dependence on the input starting point. In this study, we develop an efficient iterative algorithm for the SVI model based on a fixed-point and least-square optimizer. Furthermore, we present the convergence results in certain situations for this novel iterative algorithm. Compared with the quasi-explicit SVI method, we demonstrate the advantages of the fixed-point iterative algorithm using simulation and market data.

KEYWORDS: Iterative algorithm; FPI-SVI; SVI; Quasi-explicit SVI

## 1 Introduction

In 1999, Merrill Lynch developed the stochastic volatility inspired (SVI) parameterization model to describe the implied volatility smile appearing in the Black-Sholes option pricing formula. Owing to the profound relationship between implied variance and excellent fit with observations, the SVI model is popular in financial markets [6, 7, 14]. However, there are some limitations to the SVI model when fitting financial market data. In particular, the SVI variance smiles are convex, and do not fit several market variance smiles. In this paper, we aim to develop a efficient iterative algorithm for the SVI model.

Based on an appropriate change in variables, [8] showed that the SVI model in [6] is an appropriate solution for the implied variance in the Heston model considered in [4], where [4] established an approximate formula expressed for implied volatility functions in the Heston model. The SVI model provides a simpler expression for asymptotic implied volatility in the Heston model. [2] developed novel implied stochastic volatility models to reproduce the characteristics of the observed implied volatility smiles. Using implied volatility market data, [2] verified which stochastic volatility models are capable of reproducing the observed characteristics of implied volatility smiles (refer to [1]). [2] shared ideas with [6], and the SVI model can reproduce the characteristics of the implied volatility smiles. [9] studied the arbitrage-free SVI volatility smiles and established a large class of closed-form SVI volatility smiles (refer to [10, 11, 12, 13]). Including in particular the SVI model as an important example, [3] derived a closed-form formula for Black-Scholes implied volatility.

*Shandong University-Zhong Tai Securities Institute for Financial Studies, Shandong University, PR China, (yangsz@sdu.edu.cn). This work was supported by the National Key R&D program of China (Grant No.2018YFA0703900), National Natural Science Foundation of China (Grant No.11701330), and Young Scholars Program of Shandong University.

†Shandong University-Zhong Tai Securities Institute for Financial Studies, Shandong University, PR China, (zhangwendy@mail.sdu.edu.cn).

In most SVI optimization algorithms, an initial guess is used as a local minimizer, which leads to a strong dependence on the input starting point. It is important to find a stable and efficient algorithm for the SVI model from the practical analysis. Based on an initial estimate of two parameters in the SVI model, [15] introduced a quasi-explicit SVI method using an explicit least-squares optimizer, and thus reducing the number of SVI model parameters from five to two. Nelder–Mead was used to optimize the remaining two parameters. However, the quasi-explicit SVI method performs better based on a smart initial guess (refer to [5]).

This study focused on establishing an efficient iterative algorithm for the SVI model. The SVI model has five parameters $(a, b, \rho, m, \sigma)$ that satisfy

$$v(x) = a + b\left(\rho(x - m) + \sqrt{(x - m)^2 + \sigma^2}\right), \qquad (1.1)$$

where $x = \log K - \log F_T$, $K$ denotes the strict price in the Black-Sholes formula, $F_T$ denotes the price of the forward with maturity $T$, and $v(x)$ denotes the related implied variance. Each parameter of $(a, b, \rho, m, \sigma)$ can capture certain properties of implied variance $v(x)$. We denote by $(x_{min}, v_{min})$ the minimum point of the SVI model when $\rho^2 < 1$ and have an explicit formula with parameters $(a, b, \rho, m, \sigma)$,

$$m = x_{min} + \frac{\rho(v_{min} - a)}{b(1 - \rho^2)}, \quad \sigma = \frac{v_{min} - a}{b\sqrt{1 - \rho^2}}. \qquad (1.2)$$

This formula motivated the development of a multistep iterative algorithm. Using an approach similar to that in [15], for a given initial guess $(m_0, \sigma_0)$, we can use the least-squares optimizer to obtain $(a_0, b_0, \rho_0)$, and then use formula (1.2) to calibrate the initial guess as $(m_1, \sigma_1)$. The above method is an iterative algorithm that combines the fixed-point $(x_{min}, v_{min})$ and least-squares optimizer. We call this method the fixed-point iterative SVI (FPI-SVI) algorithm. For the case $\rho^2 = 1$, based on a simple coordinate rotation transformation, we translate the case $\rho^2 = 1$ to $\rho^2 < 1$. Furthermore, we provide a uniform FPI-SVI algorithm to deal with the cases $\rho^2 < 1$ and $\rho^2 = 1$.

The main contributions of this study are twofold:

(i). We develop a novel explicit iterative method based on the minimum point in the SVI model and the least-squares optimizer. As each step of the FPI-SVI algorithm has an explicit formula, the new algorithm is efficient. In the simulation and empirical analysis, the quasi-explicit SVI method requires almost 50 times the calculation time of the FPI-SVI algorithm with the same numbers of iterative steps.

(ii). We establish some convergence results of FPI-SVI algorithm for some situations and show that estimations of parameters $(a, b, \rho, m, \sigma)$ satisfy the least-squares optimizer and fixed-point constrained condition (1.2). In the simulation and empirical analysis, the FPI-SVI algorithm has a fast convergence rate when iterative steps are smaller than 50. The constrained condition (1.2) can improve the accuracy of the estimations of parameters $(a, b, \rho, m, \sigma)$.

The remainder of this paper is organized as follows. Section 2 introduces a novel iterative algorithm for parameters in the SVI model and establishes some convergence results for this algorithm under certain situations. Based on simulation and empirical analysis, we show some advantages of our FPI-SVI algorithm compared with the quasi-explicit SVI method in Sections 3 and 4. Finally, we conclude the study in Section 5.

## 2 FPI-SVI Algorithm

The SVI model perfectly fits several real market data. However, most optimizer algorithms of the SVI model strongly depend on the input starting point. We now show the details of the FPI-SVI algorithm used in this study. We consider the case $\rho^2 < 1$ and the minimum value of $v(x)$,

$$v_{min} = v(x_{min}) = a + b\sigma \sqrt{1-\rho^2}, \quad x_{min} = m - \frac{\rho\sigma}{\sqrt{1-\rho^2}},$$

which deduces that

$$m = x_{min} + \frac{\rho(v_{min} - a)}{b(1-\rho^2)}, \qquad (2.1)$$

and

$$\sigma = \frac{v_{min} - a}{b\sqrt{1-\rho^2}}. \qquad (2.2)$$

Now, we consider the sequences $\{x_i, v_i\}_{i=1}^N$ which are the observed points of the model (1.1). Determining the minimum point $(x_{min}, v_{min})$ of model (1.1) is important in our new FPI-SVI algorithm. In the following, we present three methods for determining the minimum point $(x_{min}, v_{min})$ when $\rho^2 < 1$.

**Remark 2.1.** *We proposed three methods for finding a better minimum point based on the observations $\{x_i, v_i\}_{i=1}^N$ for the FPI-SVI algorithm developed in this study:*

- ***Method I***: *A natural method is to use the minimum point of $\{x_i, v_i\}_{i=1}^{N}$ to estimate $(x_{min}, v_{min})$. Let $p = \arg\min_{1 \leq i \leq N} v_i$, and thus $(x_{min}, v_{min}) = (x_p, v_p)$;*

- ***Method II***: *Based on the minimum point of $\{x_i, v_i\}_{i=1}^{N}$, we use a smooth function to approximate the local property of the SVI curve and calibrate the minimum point $(x_{min}, v_{min})$ that satisfies*

$$x_{min} = x_p, \ v_{min} = v_p, \ p = \arg\min_{1 \leq i \leq N} v_i.$$

  *We take three points from observations $\{x_i, v_i\}_{i=1}^{N}$ that are close to $(x_p, v_p)$,*

$$(x_{p-1}, v_{p-1}), \ (x_p, v_p), \ (x_{p+1}, v_{p+1}).$$

  *We then use a quadratic function to fit the above three points,*

$$v(x) = \hat{c}_1 x^2 + \hat{c}_2 x + \hat{c}_3$$

  *and obtain the parameters of the quadratic function $(\hat{c}_1, \hat{c}_2, \hat{c}_3)$. Thus, we obtain the calibration minimum point, denoted by $(x_{min}, v_{min})$, and*

$$(x_{min}, v_{min}) = \left(-\frac{\hat{c}_2}{2\hat{c}_1}, \frac{4\hat{c}_1 \hat{c}_3 - \hat{c}_2^2}{4\hat{c}_1}\right).$$

- ***Method III***: *Based on the minimum point $(x_p, v_p)$ of $\{x_i, v_i\}_{i=1}^{N}$, we guess that the minimum point of the SVI model (1.1) is close to the three points*

$$(x_{p-1}, v_{p-1}), \ (x_p, v_p), \ (x_{p+1}, v_{p+1}).$$

  *We consider the set $A := \{(x, y) : \sqrt{(x - x_p)^2 + (y - y_p)^2} < r\}$, where $0 < r$ and $r$ is the maximum value such that $(x_{p-1}, v_{p-1}) \notin A$ or $(x_{p+1}, v_{p+1}) \notin A$. We can randomly choose several points (for example 10) from set A, and consider our FPI-SVI method under each given point regarded as the minimum point $(x_{min}, v_{min})$. Finally, according to the performance of the FPI-SVI algorithm, we can find a better minimum point.*

In practical analysis, we find that **Method II** is better for estimating the minimum point of the model (1.1). Thus, we consider **Method II** in Section 4. Furthermore, we can use **Method III** to find a better estimation of the minimum point of the model (1.1). However, **Method III** requires a significant amount of time compared to **Method II**.

We now consider the following two cases: $\rho^2 < 1$ and $\rho^2 = 1$. For $\rho^2 < 1$, we first establish the convergence results for the new FPI-SVI algorithm under certain situations. Subsequently, based on a simple coordinate rotation transformation, we show that one can translate the case $\rho^2 = 1$ to $\rho^2 < 1$.

## 2.1 When $\rho^2 < 1$

We denote by

$$V = (v_1, v_2, \cdots, v_N)^\top \in \mathbb{R}^{N \times 1}, \quad Y(m, \sigma) = (X_1, X_2(m, \sigma), X_3(m, \sigma)) \in \mathbb{R}^{N \times 3},$$

where

$$X_1 = (1, 1, \cdots, 1)^\top;$$
$$X_2(m, \sigma) = (x_1 - m, x_2 - m, \cdots, x_N - m)^\top;$$
$$X_3(m, \sigma) = \left( \sqrt{(x_1 - m)^2 + \sigma^2}, \sqrt{(x_2 - m)^2 + \sigma^2}, \cdots, \sqrt{(x_N - m)^2 + \sigma^2} \right)^\top.$$

We also construct an iterative algorithm based on the fixed point $(x_{min}, v_{min})$ and the quasi-explicit SVI model. We first consider the starting input point and take

$$(m_0, \sigma_0) = (x_{min}, v_{min}).$$

From the observations $(Y(m_0, \sigma_0), V)$ and denoting by $Y_0 = Y(m_0, \sigma_0)$, we introduce the following least-square optimization problem:

$$\min_{\beta} (V - Y_0 \beta)^\top (V - Y_0 \beta), \tag{2.3}$$

where $\beta = (a, b\rho, b)^\top$. The optimizer for problem (2.3) is given as follows:

$$\beta_0 = [Y_0^\top Y_0]^{-1} Y_0^\top V. \tag{2.4}$$

Thus, the estimations of $(a, b, \rho)$ are:

$$(a_0, b_0, \rho_0) = \left( \beta_0(1), \beta_0(3), \frac{\beta_0(2)}{\beta_0(3)} \right). \tag{2.5}$$

By combining equations (2.1) and (2.2), we can obtain the value of $(m, \sigma)$ at step 1:

$$m_1 = x_{min} + \frac{\rho_0(v_{min} - a_0)}{b_0(1 - \rho_0^2)}, \quad \sigma_1 = \frac{v_{min} - a_0}{b_0 \sqrt{1 - \rho_0^2}}. \tag{2.6}$$

Now, we can repeat equations (2.4), (2.5), and (2.6) from step $n$ to step $n + 1$.

$$Y_n = Y(m_n, \sigma_n); \tag{2.7}$$
$$\beta_n = [Y_n^\top Y_n]^{-1} Y_n^\top V; \tag{2.8}$$
$$(a_n, b_n, \rho_n) = \left( \beta_n(1), \beta_n(3), \frac{\beta_n(2)}{\beta_n(3)} \right); \tag{2.9}$$
$$m_{n+1} = x_{min} + \frac{\rho_n(v_{min} - a_n)}{b_n(1 - \rho_n^2)}, \quad \sigma_{n+1} = \frac{v_{min} - a_n}{b_n \sqrt{1 - \rho_n^2}}. \tag{2.10}$$

We conclude the FPI-SVI algorithm as follows:

---

**Algorithm 1:** FPI-SVI Algorithm

---

**Input:** $(m_0, \sigma_0) = (x_{min}, v_{min})$, $\{x_i, v_i\}_{i=1}^N$

**Output:** Estimations of parameters $(a, b, \rho, m, \sigma)$

---

**Initialization**: $n = 0$, $M = 50$;

Error $\delta = 1.0e - 3$;

$Y_0 = Y(m_0, \sigma_0)$;

$\beta_0 = \left[Y_0^\top Y_0\right]^{-1} Y_0^\top V$;

$L(0) = \sqrt{(V - Y_0 \beta_0)^\top (V - Y_0 \beta_0)}$;

$(a_0, b_0, \rho_0) = \left(\beta_0(1), \beta_0(3), \frac{\beta_0(2)}{\beta_0(3)}\right)$.

---

**while** $L(n) > \delta$ *or* $n \leq M$ **do**

$\quad n = n + 1$;

$\quad m_n = x_{min} + \frac{\rho_{n-1}(v_{min} - a_{n-1})}{b_{n-1}(1 - \rho_{n-1}^2)}$, $\sigma_n = \frac{v_{min} - a_{n-1}}{b_{n-1}\sqrt{1 - \rho_{n-1}^2}}$;

$\quad Y_n = Y(m_n, \sigma_n)$;

$\quad \beta_n = [Y_n^\top Y_n]^{-1} Y_n^\top V$;

$\quad L(n) = \sqrt{(V - Y_n \beta_n)^\top (V - Y_n \beta_n)}$;

$\quad (a_n, b_n, \rho_n) = \left(\beta_n(1), \beta_n(3), \frac{\beta_n(2)}{\beta_n(3)}\right)$.

**end**

---

**Remark 2.2.** *The algorithm 1 provides details of the pseudocode for the FPI-SVI algorithm. Each step of the FPI-SVI algorithm has an explicit formula. Therefore, the FPI-SVI is an efficient iterative algorithm. Compared to the quasi-explicit SVI method, we use simulation and financial market data to verify the performance of the FPI-SVI algorithm in Sections 3 and 4.*

It is theoretically challenging to directly show the convergence of the sequences $\{a_n, b_n, \rho_n, m_n, \sigma_n\}_{n=1}^\infty$. Therefore, we present situations that guarantee the convergence of $\{a_n, b_n, \rho_n, m_n, \sigma_n\}_{n=1}^\infty$. These situations are useful for practical analysis.

**Lemma 2.1.** *Let the sequences $\{a_n, b_n, \rho_n\}_{n=1}^\infty$ satisfy the following conditions:*

*(i). $0 < \underline{L}_b \leq b_n$, $|\rho_n| < L_\rho < 1$, $0 \leq a_n$, $n \geq 1$;*

*(ii). For a sufficiently small $\delta > 0$, there is a positive integer $N_0$, such that*

$$\left|a_{N_0} - a_{N_0-1}\right| < \delta;$$
$$\left|b_{N_0} - b_{N_0-1}\right| < \delta;$$
$$\left|\rho_{N_0} - \rho_{N_0-1}\right| < \delta.$$

*Then, we have that,*

$$\left|m_{N_0+1} - m_{N_0}\right| \leq L_m \delta, \quad \left|\sigma_{N_0+1} - \sigma_{N_0}\right| \leq L_\sigma \delta,$$

*where $L_m$ and $L_\sigma$ depend on $\underline{L}_b$ and $L_\rho$.*

*Proof.* First, we prove inequality $\left|m_{N_0+1} - m_{N_0}\right| \leq L_m \delta$. Note that

$$m_{N_0+1} = x_{min} + \frac{\rho_{N_0}(v_{min} - a_{N_0})}{b_{N_0}(1 - \rho_{N_0}^2)}, \quad m_{N_0} = x_{min} + \frac{\rho_{N_0-1}(v_{min} - a_{N_0-1})}{b_{N_0-1}(1 - \rho_{N_0-1}^2)},$$

which deduces that

$$\begin{aligned}
\left|m_{N_0+1} - m_{N_0}\right| &\leq \left|\frac{\rho_{N_0}(v_{min} - a_{N_0})}{b_{N_0}(1 - \rho_{N_0}^2)} - \frac{\rho_{N_0-1}(v_{min} - a_{N_0-1})}{b_{N_0-1}(1 - \rho_{N_0-1}^2)}\right| \\
&\leq \left|\frac{\rho_{N_0}(v_{min} - a_{N_0})}{b_{N_0}(1 - \rho_{N_0}^2)} - \frac{\rho_{N_0-1}(v_{min} - a_{N_0})}{b_{N_0}(1 - \rho_{N_0-1}^2)}\right| \\
&+ \left|\frac{\rho_{N_0-1}(v_{min} - a_{N_0})}{b_{N_0}(1 - \rho_{N_0-1}^2)} - \frac{\rho_{N_0-1}(v_{min} - a_{N_0-1})}{b_{N_0}(1 - \rho_{N_0-1}^2)}\right| \\
&+ \left|\frac{\rho_{N_0-1}(v_{min} - a_{N_0-1})}{b_{N_0}(1 - \rho_{N_0-1}^2)} - \frac{\rho_{N_0-1}(v_{min} - a_{N_0-1})}{b_{N_0-1}(1 - \rho_{N_0-1}^2)}\right| \\
&\leq \frac{(1 + L_\rho^2)v_{min}}{\underline{L}_b(1 - L_\rho^2)^2}\delta + \frac{L_\rho}{\underline{L}_b(1 - L_\rho^2)}\delta + \frac{L_\rho v_{min}}{\underline{L}_b^2(1 - L_\rho^2)}\delta \\
&= L_m \delta,
\end{aligned}$$

where

$$L_m = \frac{\underline{L}_b(1 + L_\rho^2)v_{min} + \underline{L}_b L_\rho(1 - L_\rho^2) + L_\rho(1 - L_\rho^2)v_{min}}{\underline{L}_b^2(1 - L_\rho^2)^2}.$$

Similarly, we have $\left|\sigma_{N_0+1} - \sigma_{N_0}\right| \leq L_\sigma \delta$, where

$$L_\sigma = \frac{L_\rho \underline{L}_b v_{min} + \underline{L}_b(1 - L_\rho^2) + (1 - L_\rho^2)v_{min}}{\underline{L}_b^2(1 - L_\rho^2)^{\frac{3}{2}}}.$$

□

For a given positive integer $N_0 > 0$, we introduce the following notations used in Lemma 2.2:

$L_{0,N_0}(1) = N\left|\beta_{N_0}(2)\right|L_m + N\left|\beta_{N_0}(3)\right|(L_m + L_\sigma);$

$L_{0,N_0}(2) = \left|X_1^\top[V - Y(m_{N_0}, \sigma_{N_0})\beta_{N_0}^1]\right|L_m + \left|X_1^\top X_2(m_{N_0}, \sigma_{N_0})\beta_{N_0}(3)\right|(L_m + L_\sigma) + \left|\beta_{N_0}(2)\right| + \left|\beta_{N_0}(3)\right|;$

$L_{0,N_0}(3) = \left|X_1^\top[V - Y(m_{N_0}, \sigma_{N_0})\beta_{N_0}^2]\right|(L_m + L_\sigma) + \left|X_1^\top X_3(m_{N_0}, \sigma_{N_0})\beta_{N_0}(2)\right|L_m + \left|\beta_{N_0}(2)\right| + \left|\beta_{N_0}(3)\right|,$

where $\beta_{N_0}^1 = (\beta_{N_0}(1), 2\beta_{N_0}(2), \beta_{N_0}(3))^\top$ and $\beta_{N_0}^2 = (\beta_{N_0}(1), \beta_{N_0}(2), 2\beta_{N_0}(3))^\top$.

**Lemma 2.2.** *Let the sequences $\{a_n, b_n, \rho_n\}_{n=1}^{\infty}$ satisfy the following conditions:*

*(i). $0 < \underline{L}_b \leq b_n$, $|\rho_n| < L_\rho < 1$, $0 \leq a_n$, $n \geq 1$;*

*(ii). For a sufficiently small $\delta > 0$, there is a positive integer $N_0$, such that*

$$\left|a_{N_0} - a_{N_0-1}\right| < \delta;$$
$$\left|b_{N_0} - b_{N_0-1}\right| < \delta;$$
$$\left|\rho_{N_0} - \rho_{N_0-1}\right| < \delta;$$

*(iii). Absolute value of each element of the vector*

$$[Y_{N_0}^\top Y_{N_0}]^{-1} L_{0,N_0}$$

*is smaller than $(1-\alpha)L$, where $0 < L < 1$, $0 < \alpha < 1$, $L_{0,N_0} = (L_{0,N_0}(1), L_{0,N_0}(2), L_{0,N_0}(3))^\top$.*

*Then, we have that,*

$$\left|a_{N_0+1} - a_{N_0}\right| < L\delta;$$
$$\left|b_{N_0+1} - b_{N_0}\right| < L\delta;$$
$$\left|\rho_{N_0+1} - \rho_{N_0}\right| < \frac{2L}{\underline{L}_b}\delta.$$

*Proof.* Note that

$$V = (v_1, v_2, \cdots, v_N)^\top \in \mathbb{R}^{N\times 1}, \quad Y(m_{N_0+1}, \sigma_{N_0+1}) = (X_1, X_2(m_{N_0+1}, \sigma_{N_0+1}), X_3(m_{N_0+1}, \sigma_{N_0+1})) \in \mathbb{R}^{N\times 3},$$

where

$$X_1 = (1, 1, \cdots, 1)^\top;$$
$$X_2(m_{N_0+1}, \sigma_{N_0+1}) = (x_1 - m_{N_0+1}, \cdots, x_N - m_{N_0+1})^\top;$$
$$X_3(m_{N_0+1}, \sigma_{N_0+1}) = \left(\sqrt{(x_1 - m_{N_0+1})^2 + \sigma_{N_0+1}^2}, \cdots, \sqrt{(x_N - m_{N_0+1})^2 + \sigma_{N_0+1}^2}\right)^\top.$$

The following two formulas are considered:

$$\beta_{N_0} = [Y_{N_0}^\top Y_{N_0}]^{-1} Y_{N_0}^\top V, \quad \beta_{N_0+1} = [Y_{N_0+1}^\top Y_{N_0+1}]^{-1} Y_{N_0+1}^\top V,$$

where $\beta_{N_0} = (a_{N_0}, b_{N_0}\rho_{N_0}, b_{N_0})^\top$ and $\beta_{N_0+1} = (a_{N_0+1}, b_{N_0+1}\rho_{N_0+1}, b_{N_0+1})^\top$. Following that,

$$\beta_{N_0+1} - \beta_{N_0} = [Y_{N_0+1}^\top Y_{N_0+1}]^{-1} Y_{N_0+1}^\top V - [Y_{N_0}^\top Y_{N_0}]^{-1} Y_{N_0}^\top V. \tag{2.11}$$

Let $\Delta = (\Delta_0, \Delta_1, \Delta_2)$, where

$$\Delta_0 = (0, 0, \cdots, 0)^\top;$$
$$\Delta_1 = (\underline{\delta}_1, \underline{\delta}_2, \cdots, \underline{\delta}_N)^\top;$$
$$\Delta_2 = (\overline{\delta}_1, \overline{\delta}_2, \cdots, \overline{\delta}_N)^\top,$$

and $\underline{\delta}_i = (x_i - m_{N_0+1}) - (x_i - m_{N_0})$, $\overline{\delta}_i = \sqrt{(x_i - m_{N_0+1})^2 + \sigma_{N_0+1}^2} - \sqrt{(x_i - m_{N_0})^2 + \sigma_{N_0}^2}$, $1 \leq i \leq N$.
For any given $i$, by Lemma 2.1, we have that

$$\left|\underline{\delta}_i\right| = \left|m_{N_0+1} - m_{N_0}\right| < L_m \delta,$$

and

$$\begin{aligned}
\left|\overline{\delta}_i\right| &= \left|\sqrt{(x_i - m_{N_0+1})^2 + \sigma_{N_0+1}^2} - \sqrt{(x_i - m_{N_0})^2 + \sigma_{N_0}^2}\right| \\
&\leq \left|\sqrt{(x_i - m_{N_0+1})^2 + \sigma_{N_0+1}^2} - \sqrt{(x_i - m_{N_0})^2 + \sigma_{N_0+1}^2}\right| \\
&\quad + \left|\sqrt{(x_i - m_{N_0})^2 + \sigma_{N_0+1}^2} - \sqrt{(x_i - m_{N_0})^2 + \sigma_{N_0}^2}\right| \\
&\leq \frac{\left|(x_i - m_{N_0+1})^2 - (x_i - m_{N_0})^2\right|}{\left|x_i - m_{N_0+1}\right| + \left|x_i - m_{N_0}\right|} + \frac{\left|\sigma_{N_0+1}^2 - \sigma_{N_0}^2\right|}{\sigma_{N_0+1} + \sigma_{N_0}} \\
&\leq (L_m + L_\sigma)\delta,
\end{aligned}$$

and thus

$$\begin{aligned}
\left|\underline{\delta}_i\right| &\leq L_m \delta; \\
\left|\overline{\delta}_i\right| &\leq (L_m + L_\sigma)\delta; \\
Y_{N_0+1} &= \Delta + Y_{N_0}.
\end{aligned}$$

We first consider the right part of (2.11):

$$\begin{aligned}
&[Y_{N_0+1}^\top Y_{N_0+1}]^{-1} Y_{N_0+1}^\top V - [Y_{N_0}^\top Y_{N_0}]^{-1} Y_{N_0}^\top V \\
&= [Y_{N_0}^\top Y_{N_0} + Y_{N_0}^\top \Delta + \Delta^\top Y_{N_0} + \Delta^\top \Delta]^{-1} [Y_{N_0}^\top + \Delta^\top] V - [Y_{N_0}^\top Y_{N_0}]^{-1} Y_{N_0}^\top V \\
&= [Y_{N_0}^\top Y_{N_0}]^{-1} \left[ Y_{N_0}^\top Y_{N_0} [Y_{N_0}^\top Y_{N_0} + Y_{N_0}^\top \Delta + \Delta^\top Y_{N_0} + \Delta^\top \Delta]^{-1} [Y_{N_0}^\top + \Delta^\top] V - Y_{N_0}^\top V \right] \\
&= [Y_{N_0}^\top Y_{N_0}]^{-1} \left[ \left[ I + [Y_{N_0}^\top \Delta + \Delta^\top Y_{N_0} + \Delta^\top \Delta][Y_{N_0}^\top Y_{N_0}]^{-1} \right]^{-1} - I \right] Y_{N_0}^\top V \\
&\quad + [Y_{N_0}^\top Y_{N_0}]^{-1} \left[ \left[ I + [Y_{N_0}^\top \Delta + \Delta^\top Y_{N_0} + \Delta^\top \Delta][Y_{N_0}^\top Y_{N_0}]^{-1} \right]^{-1} \Delta^\top V \right].
\end{aligned}$$

From

$$\begin{aligned}
A &= \left[ I + B[Y_{N_0}^\top Y_{N_0}]^{-1} \right]^{-1}; \\
B &= Y_{N_0}^\top \Delta + \Delta^\top Y_{N_0} + \Delta^\top \Delta,
\end{aligned}$$

we obtain

$$\begin{aligned}
I &= A \left[ I + B[Y_{N_0}^\top Y_{N_0}]^{-1} \right] \\
&= A + AB[Y_{N_0}^\top Y_{N_0}]^{-1},
\end{aligned}$$

and
$$A = I - AB[Y_{N_0}^\top Y_{N_0}]^{-1}.$$

Furthermore, from (2.11), it follows that
$$\beta_{N_0+1} - \beta_{N_0} = [Y_{N_0}^\top Y_{N_0}]^{-1} A \left[\Delta^\top V - B\beta_{N_0}\right],$$

and $\Delta^\top V - B\beta_{N_0} = (E_1, E_2, E_3)^\top$, where:

$$E_1 = -\sum_{i=1}^{N} \left(\underline{\delta}_i \beta_{N_0}(2) + \overline{\delta}_i \beta_{N_0}(3)\right);$$

$$E_2 = \sum_{i=1}^{N} \left(\underline{\delta}_i v_i - \underline{\delta}_i \beta_{N_0}(1) - \left[2\underline{\delta}_i(x_i - m_{N_0}) + \underline{\delta}_i^2\right]\beta_{N_0}(2)\right.$$
$$\left. - \left[\overline{\delta}_i(x_i - m_{N_0}) + \underline{\delta}_i \sqrt{(x_i - m_{N_0})^2 + \sigma_{N_0}^2} + \underline{\delta}_i \overline{\delta}_i\right]\beta_{N_0}(3)\right);$$

$$E_3 = \sum_{i=1}^{N} \left(\overline{\delta}_i v_i - \overline{\delta}_i \beta_{N_0}(1) - \left[\overline{\delta}_i(x_i - m_{N_0}) + \underline{\delta}_i \sqrt{(x_i - m_{N_0})^2 + \sigma_{N_0}^2} + \underline{\delta}_i \overline{\delta}_i\right]\beta_{N_0}(2)\right.$$
$$\left. - \left[2\overline{\delta}_i \sqrt{(x_i - m_{N_0})^2 + \sigma_{N_0}^2} + \overline{\delta}_i^2\right]\beta_{N_0}(3)\right).$$

From the inequalities of $\underline{\delta}_i, \overline{\delta}_i,\ 1 \leq i \leq N$, and sufficiently small $\delta$, we have
$$|E_i| \leq L_{0,N_0}(i)\delta, \quad i = 1, 2, 3,$$

where

$$L_{0,N_0}(1) = N\left|\beta_{N_0}(2)\right| L_m + N\left|\beta_{N_0}(3)\right|(L_m + L_\sigma);$$
$$L_{0,N_0}(2) = \left|X_1^\top[V - Y(m_{N_0}, \sigma_{N_0})\beta_{N_0}^1]\right| L_m + \left|X_1^\top X_2(m_{N_0}, \sigma_{N_0})\beta_{N_0}(3)\right|(L_m + L_\sigma) + \left|\beta_{N_0}(2)\right| + \left|\beta_{N_0}(3)\right|;$$
$$L_{0,N_0}(3) = \left|X_1^\top[V - Y(m_{N_0}, \sigma_{N_0})\beta_{N_0}^2]\right|(L_m + L_\sigma) + \left|X_1^\top X_3(m_{N_0}, \sigma_{N_0})\beta_{N_0}(2)\right| L_m + \left|\beta_{N_0}(2)\right| + \left|\beta_{N_0}(3)\right|,$$

and $\beta_{N_0}^1 = (\beta_{N_0}(1), 2\beta_{N_0}(2), \beta_{N_0}(3))^\top$, $\beta_{N_0}^2 = (\beta_{N_0}(1), \beta_{N_0}(2), 2\beta_{N_0}(3))^\top$.

For a sufficiently small $\delta > 0$, it is convenient to demonstrate that the absolute value of each element of the vector
$$A\left[\Delta^\top V - B\beta_{N_0}\right],$$

is smaller than that of vector
$$\frac{\delta}{1-\alpha} L_{0,N_0}, \quad L_{0,N_0} = (L_{0,N_0}(1), L_{0,N_0}(2), L_{0,N_0}(3))^\top,$$

where $0 < \alpha < 1$ is a given constant. Then, from condition (iii), we have
$$\left|\beta_{N_0+1}(i) - \beta_{N_0}(i)\right| \leq L\delta,\ L < 1,\ i = 1, 2, 3,$$

which deduces that

$$|a_{N_0+1} - a_{N_0}| < L\delta;$$
$$|b_{N_0+1}\rho_{N_0+1} - b_{N_0}\rho_{N_0}| < L\delta;$$
$$|b_{N_0+1} - b_{N_0}| < L\delta.$$

From the inequality $|b_{N_0+1}\rho_{N_0+1} - b_{N_0}\rho_{N_0}| < L\delta$, following that

$$|b_{N_0+1}\rho_{N_0+1} - b_{N_0}\rho_{N_0}|$$
$$= |b_{N_0+1}\rho_{N_0+1} - b_{N_0+1}\rho_{N_0} + b_{N_0+1}\rho_{N_0} - b_{N_0}\rho_{N_0}|$$
$$\geq |b_{N_0+1}\rho_{N_0+1} - b_{N_0+1}\rho_{N_0}| - |b_{N_0+1}\rho_{N_0} - b_{N_0}\rho_{N_0}|$$

and thus

$$|\rho_{N_0+1} - \rho_{N_0}| \leq \frac{2L}{\underline{L}_b}\delta.$$

This completes the proof. $\square$

Based on Lemmas 2.1 and 2.2, we present the main results.

**Theorem 2.1.** *Let conditions (i) and (ii) of Lemma 2.2 hold; condition (iii) in Lemma 2.2 is independent of $N_0$, i.e,*

*(iii').*

$$\sup_{n \geq N_0} \left|[Y_n^\top Y_n]^{-1} L_{0,n}(i)\right| < (1-\alpha)L, \ i = 1, 2, 3,$$

*where $0 < \alpha < 1$, $[Y_n^\top Y_n]^{-1} L_{0,n}(i)$ is the i-th element of the vector $[Y_n^\top Y_n]^{-1} L_{0,n}$, $L_{0,n}$ is given in Lemma 2.2, $0 < L < 1$, and $2L < \underline{L}_b$.*

*Then, we have that*

$$|a_n - a_{n-1}| < \left(L \vee \frac{2L}{\underline{L}_b}\right)^{n-N_0} \delta;$$
$$|b_n - b_{n-1}| < \left(L \vee \frac{2L}{\underline{L}_b}\right)^{n-N_0} \delta;$$
$$|\rho_n - \rho_{n-1}| < \left(L \vee \frac{2L}{\underline{L}_b}\right)^{n-N_0} \delta,$$

*and the sequences $\{a_n, b_n, \rho_n\}_{n \geq N_0}$ converge as $n \to \infty$.*

*Proof.* By applying Lemma 2.2, for $N_0$, we have

$$\left|a_{N_0+1} - a_{N_0}\right| < L\delta;$$
$$\left|b_{N_0+1} - b_{N_0}\right| < L\delta;$$
$$\left|\rho_{N_0+1} - \rho_{N_0}\right| < \frac{2L}{\underline{L}_b}\delta.$$

Note that $L < 1$ and $2L < \underline{L}_b$; then, we can apply Lemma 2.2 to the case $N_0 + 1$, and obtain

$$\left|a_{N_0+2} - a_{N_0+1}\right| < \left(L \vee \frac{2L}{\underline{L}_b}\right)^2 \delta;$$
$$\left|b_{N_0+2} - b_{N_0+1}\right| < \left(L \vee \frac{2L}{\underline{L}_b}\right)^2 \delta;$$
$$\left|\rho_{N_0+2} - \rho_{N_0+1}\right| < \left(L \vee \frac{2L}{\underline{L}_b}\right)^2 \delta.$$

We complete the proof using the induction method. □

**Remark 2.3.** *Theorem 2.1 shows that when the difference in sequences $\{a_n, b_n, \rho_n\}_{n\geq 1}$ at some index $N_0$ is sufficiently small and the observations of model (1.1) satisfy bounded condition (iii'), the sequences $\{a_n, b_n, \rho_n\}_{n\geq 1}$ converge as $n \to \infty$. Although the conditions in Theorem 2.1 are rather complicated, we can easily verify these conditions in a practical analysis.*

Based on Theorem 2.1, we introduce the following convergence results:

**Theorem 2.2.** *Let the sequences $\{a_n, b_n, \rho_n\}_{n=1}^{\infty}$ satisfy the following conditions:*

*(i). $\{a_n, b_n, \rho_n\}_{n=1}^{\infty}$ are bounded and $\sup_{n\geq 1} |\rho_n| < 1$.*

*(ii). There exists a positive integer $N_0$ such that when $n > N_0$, $\{a_n\}_{n>N_0}$ and $\{b_n\}_{n>N_0}$ increase with n, and $\{\rho_n\}_{n>N_0}$ decreases with n.*

*or*

*(iii). A positive integer $N_0$ exists such that when $n > N_0$, $\{a_n\}_{n>N_0}$ and $\{b_n\}_{n>N_0}$ decrease with n and $\{\rho_n\}_{n>N_0}$ increases with n.*

*or*

*(iv). Sequences $\{a_n, b_n, \rho_n\}_{n\geq 1}$ converge as $n \to \infty$.*

*Then, we have $\{m_n, \sigma_n\}_{n>N_0}$ as decreasing sequences with n under conditions (i) and (ii), and $\{m_n, \sigma_n\}_{n>N_0}$ as increasing sequences with n under conditions (i) and (iii). We have limits $(a^*, b^*, \rho^*, m^*, \sigma^*)$ for sequences $\{a_n, b_n, \rho_n, m_n, \sigma_n\}_{n\geq 1}$ under conditions (i) and (ii), or (iii), or (iv).*

*Proof.* It is easy to prove that $\{m_n, \sigma_n\}_{n>N_0}$ are decreasing sequences with *n* under conditions (i) and (ii). Similarly, we can show that $\{m_n, \sigma_n\}_{n>N_0}$ are increasing sequences with *n* under conditions (i) and (iii).

For any given $n > N_0$, from Formula (2.10), it follows that

$$m_{n+1} = x_{min} + \frac{\rho_n(v_{min} - a_n)}{b_n(1 - \rho_n^2)}, \quad \sigma_{n+1} = \frac{v_{min} - a_n}{b_n \sqrt{1 - \rho_n^2}}.$$

According to conditions (i) and (ii), $\{m_{n+1}, \sigma_{n+1}\}_{n > N_0}$ decreases with $n$. Therefore, we set

$$(a^*, b^*, \rho^*, m^*, \sigma^*) = \lim_{n \to \infty}(a_n, b_n, \rho_n, m_n, \sigma_n),$$

which satisfy

$$m^* = x_{min} + \frac{\rho^*(v_{min} - a^*)}{b^*(1 - \rho^{*2})}, \quad \sigma^* = \frac{v_{min} - a^*}{b^* \sqrt{1 - \rho^{*2}}}. \tag{2.12}$$

Similarly, we can establish formulas (2.12) for $(a^*, b^*, \rho^*, m^*, \sigma^*)$ under conditions (i) and (iii) or (iv). Furthermore, we have that $(a^*, b^*, \rho^*, m^*, \sigma^*)$ are the optimal parameters of the least-squares optimizer

$$\min_{\beta = (a,b,\rho)^\top} (V - Y(m,\sigma)\beta)^\top (V - Y(m,\sigma)\beta) \tag{2.13}$$

with the minimum-point constraint condition (2.12). $\square$

**Remark 2.4.** *Based on the results of Theorem 2.2, when some monotonic conditions of $\{a_n, b_n, \rho_n\}$ are satisfied, our FPI-SVI algorithm can obtain an optimal estimation for parameters $(a, b, \rho, m, \sigma)$ for the least-squares optimizer (2.13) under constraint (2.12) at the minimum point $(x_{min}, v_{min})$. The advantage of FPI-SVI algorithm is that we can use the constraint condition (2.12) to improve the accuracy of the least-squares estimations of (2.13).*

## 2.2 When $\rho^2 = 1$

Now, we consider the case $\rho^2 = 1$ which is an important situation in real financial market. Note that, when the curve is rotated along the origin $(0, 0)$, the properties of the curve will not change. Let $\rho = -1$. It is easy to demonstrate that the SVI model (1.1) takes the minimum point $(+\infty, a)$, and $v(x)$ decreases with $x \in \mathbb{R}$. Next, we show how to rotate the curve along the origin $(0, 0)$, and translate the case $\rho = -1$ to $\rho^2 < 1$. Let $(x, v)$ be a point in

$$v(x) = a + b(-(x - m) + \sqrt{(x - m)^2 + \sigma^2}).$$

Let $0 < \theta < \frac{\pi}{2}$. After rotating the curve, the new coordinate of the point $(x, v)$ is,

$$x' = x \cos \theta - v \sin \theta,$$
$$v' = x \sin \theta + v \cos \theta.$$

By complicated calculation, we can show that $(x', v')$ satisfies the following SVI model

$$v'(x') = a' + b'(\rho'(x' - m') + \sqrt{(x' - m')^2 + \sigma'^2}),$$

where
$$\begin{cases} a' = \dfrac{a}{\cos\theta} - a_0, \\ a_0 = m_0 \cos^4\theta \tan\theta(-\dfrac{\tan\theta}{b} + 2 + \dfrac{\rho'}{\cos^2\theta b}), \\ b' = \dfrac{b}{(1 + 2b\tan\theta)\cos^2\theta}, \\ \rho' = (\dfrac{\tan\theta}{b} + \tan^2\theta - 1)\cos^2\theta, \\ m' = \dfrac{\cos^2\theta m_0 - \dfrac{a_0}{b'}}{(\dfrac{\tan\theta}{b} + \tan^2\theta - 1)\cos^2\theta}, \\ m_0 = \dfrac{a\tan\theta}{\cos\theta} - \dfrac{m}{\cos\theta}, \\ \sigma'^2 = \dfrac{\sigma^2 b}{b'} - \dfrac{2a_0^2}{b'^2} + \dfrac{2a_0 \cos^2\theta m_0}{b'}. \end{cases}$$

Let $\rho'^2 < 1$, which deduces that

$$\left[(\dfrac{\tan\theta}{b} + \tan^2\theta - 1)\cos^2\theta\right]^2 < 1$$

and thus

$$(\dfrac{\tan\theta}{b} + 2\tan^2\theta)(\dfrac{\tan\theta}{b} - 2) < 0.$$

Note that $b > 0$, thus $\theta$ should satisfy

$$0 < \theta < \arctan(2b). \tag{2.14}$$

In the following, we consider a simple example of the SVI model with parameters $(a, b, \rho, m, \sigma) = (0.5, 0.5, -1, -0.3, 0.5)$.

Figure 1: The curve of SVI under $(a, b, \rho, m, \sigma) = (0.5, 0.5, -1, -0.3, 0.5)$ and the curve after rotation.

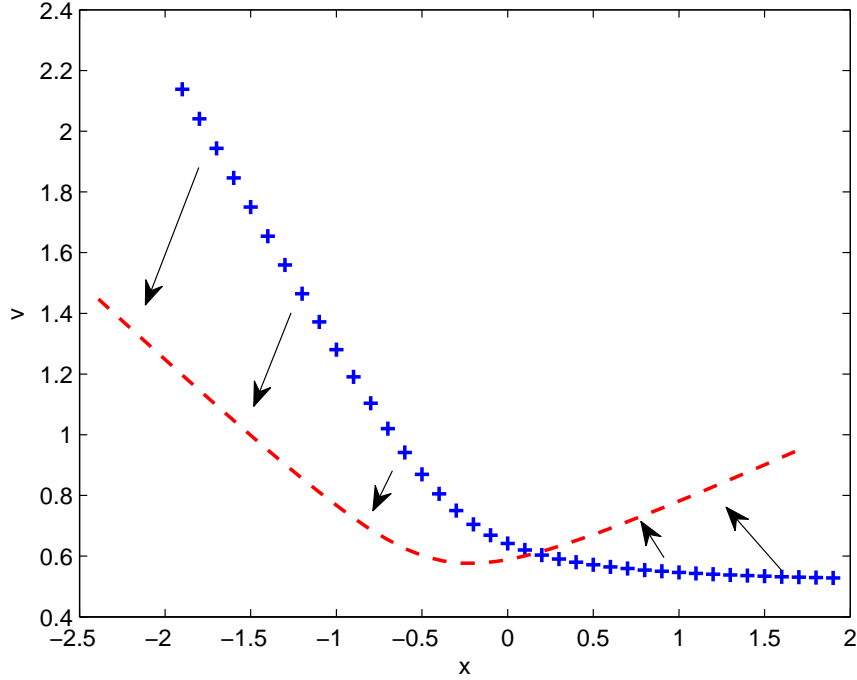

In Figure 1, we rotate the blue curve which is an SVI curve with parameters $(a, b, \rho, m, \sigma) = (0.5, 0.5, -1, -0.3, 0.5)$ to a red curve with $\theta = \frac{\pi}{12}$ which satisfies condition (2.14). We can now use the method developed for the case $\rho^2 < 1$ to estimate the parameters of the red curve. We rotate the fitting curve back, which is used to fit the blue curve. We now introduce two error indices to verify our FPI-SVI algorithm. Let $\hat{V} = (\hat{v}_1, \hat{v}_2, \cdots, \hat{v}_N)^\top$ be the points in the fitting curve. The root average squared error (RASE) and root maximum squared error (RMSE) are defined as:

$$\text{RASE} = \sqrt{\frac{(V - \hat{V})^\top (V - \hat{V})}{N}}, \quad \text{RMSE} = \sqrt{\max_{1 \leq i \leq N}(v_i - \hat{v}_i)^2}.$$

In Figure 2, we show that the RASE of the FPI-SVI algorithm is 1.2642e-04, and the RMSE is 1.9806e-04 after 50 steps, which demonstrate that our FPI-SVI algorithm is useful for the case $\rho^2 = 1$.

Figure 2: FPI-SVI algorithm: the left picture shows the estimations results of parameters in model (1.1), and the right picture shows the RASE of estimations of *v* in model (1.1).

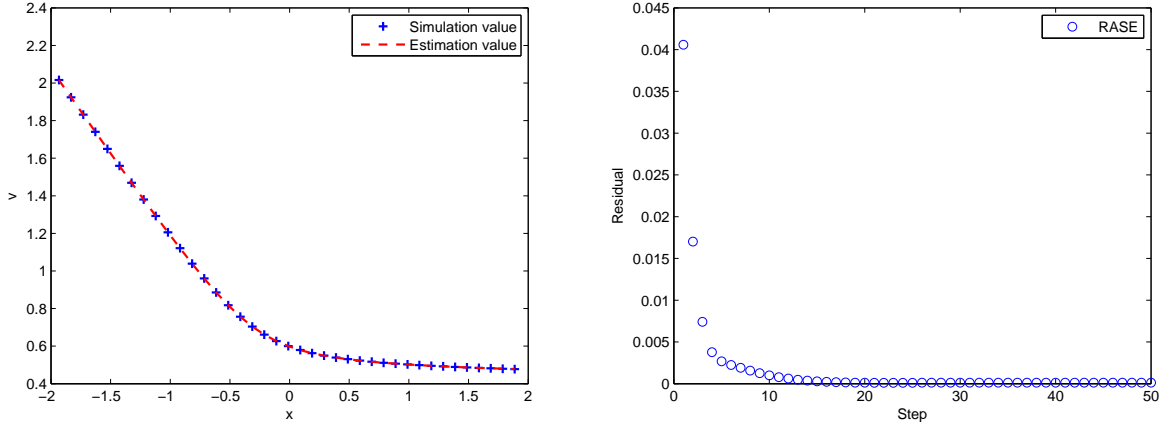

**Remark 2.5.** *In this part, we show a simple method to translate the case $\rho = -1$ to $\rho^2 < 1$ via rotating the SVI curve along the origin $(0,0)$. However, it is not possible to distinguish in advance whether the SVI model will be approximated by $\rho^2 = 1$ or $\rho^2 < 1$ in real market. These results motivate us to develop a uniform FPI-SVI algorithm to deal with the cases $\rho^2 = 1$ and $\rho^2 < 1$, such that the algorithm does not depend on the minimum value $(x_{min}, v_{min})$ of SVI model. We investigate the uniform FPI-SVI algorithm in the following subsection.*

## 2.3 A uniform FPI-SVI algorithm for $\rho^2 \leq 1$

Note that, the FPI-SVI algorithm depends on the minimum value point $(x_{min}, v_{min})$ of SVI model in Section 2.1. When $\rho^2 = 1$, we need to transform it into the case $\rho^2 < 1$ by rotating the SVI curve. In this part, we investigate a method to deal with the cases $\rho^2 < 1$ and $\rho^2 = 1$ uniformly. In the following, we show the details of the algorithm. Based on a fixed observation $\{x, v, v_x\}$, where $v_x$ is the first derivative of $v(x)$ on $x$, it follows that

$$\begin{cases} v = a + b(\rho(x - m) + \sqrt{(x-m)^2 + \sigma^2}), \\ v_x = b\rho + \dfrac{b(x-m)}{\sqrt{(x-m)^2 + \sigma^2}}, \end{cases}$$

and thus

$$\begin{cases} \dfrac{v - a}{b} = \sigma\left(\rho \dfrac{x-m}{\sigma} + \sqrt{\left(\dfrac{x-m}{\sigma}\right)^2 + 1}\right), \\ \dfrac{x-m}{\sigma} = \dfrac{v_x - b\rho}{b}\sqrt{\left(\dfrac{x-m}{\sigma}\right)^2 + 1}. \end{cases} \tag{2.15}$$

From equation (2.15), we have the following representations for *m* and $\sigma$.

**Lemma 2.3.** *Let $\{x, v, v_x\}$ be the observations from SVI curve. We have*

$$m = x - \frac{(v-a)(v_x - b\rho)}{b\rho v_x + b^2(1-\rho^2)},$$
$$\sigma = \frac{(v-a)\sqrt{b^2 - (v_x - b\rho)^2}}{b\rho v_x + b^2(1-\rho^2)}.$$
(2.16)

*Proof.* Now, we derive the explicit formulas for $\sigma$ and $m$, respectively. By the second equation of (2.15), it follows that

$$\frac{x-m}{\sigma} = \frac{v_x - b\rho}{\sqrt{b^2 - (v_x - b\rho)^2}}.$$
(2.17)

Combining equation (2.17) and the first equation of (2.15), we have

$$\frac{v-a}{b} = \frac{\rho v_x + b(1-\rho^2)}{\sqrt{b^2 - (v_x - b\rho)^2}} \sigma$$

and thus

$$\sigma = \frac{(v-a)\sqrt{b^2 - (v_x - b\rho)^2}}{b\rho v_x + b^2(1-\rho^2)}.$$
(2.18)

Then, from equations (2.17) and (2.18), it follows that

$$m = x - \frac{(v-a)(v_x - b\rho)}{b\rho v_x + b^2(1-\rho^2)}.$$
(2.19)

This completes the proof. □

**Remark 2.6.** *When the observation point is the minimum value of $v(x)$, it implies that $v_x = 0$. Then, the explicit formulas (2.16) reduce to formulas (2.1) and (2.2).*

**Remark 2.7.** *Applying Lemma 2.3, we can show the FPI-SVI algorithm following the steps given in Section 2.1. Similar with Remark 2.1, we show how to estimate $v_x$ based on the observations $\{x_i, v_i\}_{i=1}^N$.*

- ***Method I'***: *A natural method is to use the differential formula to estimate $v_x$: $v_{x,i} = \frac{v_{i+1} - v_{i-1}}{x_{i+1} - x_{i-1}}, 2 \leq i \leq N-1$. As for $v_{x,1}$ and $v_{x,N}$, we let $v_{x,1} = v_{x,2}$ and $v_{x,N} = v_{x,N-1}$.*

- ***Method II'***: *Based on a fixed observation $(x_j, v_j)$, we use a smooth function to approximate the local property of the SVI curve. We take three points from observations $\{x_i, v_i\}_{i=1}^N$. Then, we consider the following three points: $(x_{j-1}, v_{j-1})$, $(x_j, v_j)$, $(x_{j+1}, v_{j+1})$. We use a quadratic function to fit the above three points,*

$$v(x) = \hat{h}_1 x^2 + \hat{h}_2 x + \hat{h}_3$$

*and obtain the parameters of the quadratic function $(\hat{h}_1, \hat{h}_2, \hat{h}_3)$. Thus, we get the derivative function equation of $v(x)$: $v_x = 2\hat{h}_1 x + \hat{h}_2$. Then we get an initial guess point $(x_j, v_j, 2\hat{h}_1 x_j + \hat{h}_2)$.*

# 3 Simulation analysis

In this section, we first use some true parameters of the SVI model (1.1) to verify the advantages of the FPI-SVI algorithm.

We perform simulations and compare our FPI-SVI algorithm with the quasi-explicit SVI method. Considering the following true parameters of the SVI model (1.1):

$$(a, b, \rho, m, \sigma) = (0.5, 0.5, -0.5, -0.3, 0.5). \tag{3.1}$$

We take $x_i = -1.9 + 0.1(i-1)$, and $v_i = a + b\left(\rho(x_i - m) + \sqrt{(x_i - m)^2 + \sigma^2}\right)$, $1 \leq i \leq 39$. Applying the FPI-SVI algorithm 1, we perform a multistep iteration FPI-SVI algorithm with $K = 50$ and ignore the error threshold $\delta$.

First, we present the convergence property of sequences $\{a_n, b_n, \rho_n, m_n, \sigma_n\}_{n \geq 1}$ in Figure 3. In Figure 3, the values of sequences $\{a_n, b_n, \rho_n\}_{n \geq 1}$ satisfy

$$|a_{11} - a_{10}| \leq 0.01;$$
$$|b_{11} - b_{10}| \leq 0.01;$$
$$|\rho_{11} - \rho_{10}| \leq 0.01,$$

and the values of the sequences $\{m_n, \sigma_n\}_{n \geq 1}$ satisfy

$$|m_{11} - m_{10}| \leq 0.01;$$
$$|\sigma_{11} - \sigma_{10}| \leq 0.01.$$

Furthermore, the sequences $\{a_n, b_n, \rho_n, m_n, \sigma_n\}_{n \geq 1}$ converge to the limits after the 20-th iterative step, and the limits are $(0.5000, 0.5000, -0.5000, -0.3000, 0.5000)$ which are almost identical to the true values of the parameters $(a, b, \rho, m, \sigma)$. The calculation results verify the main results of Theorem 2.1. See Table 1.

Figure 3: FPI-SVI algorithm: values of sequences $\{a_n, b_n, \rho_n, m_n, \sigma_n\}_{n=1}^{50}$ along with the true values of parameters $(a, b, \rho, m, \sigma) = (0.5, 0.5, -0.5, -0.3, 0.5)$.

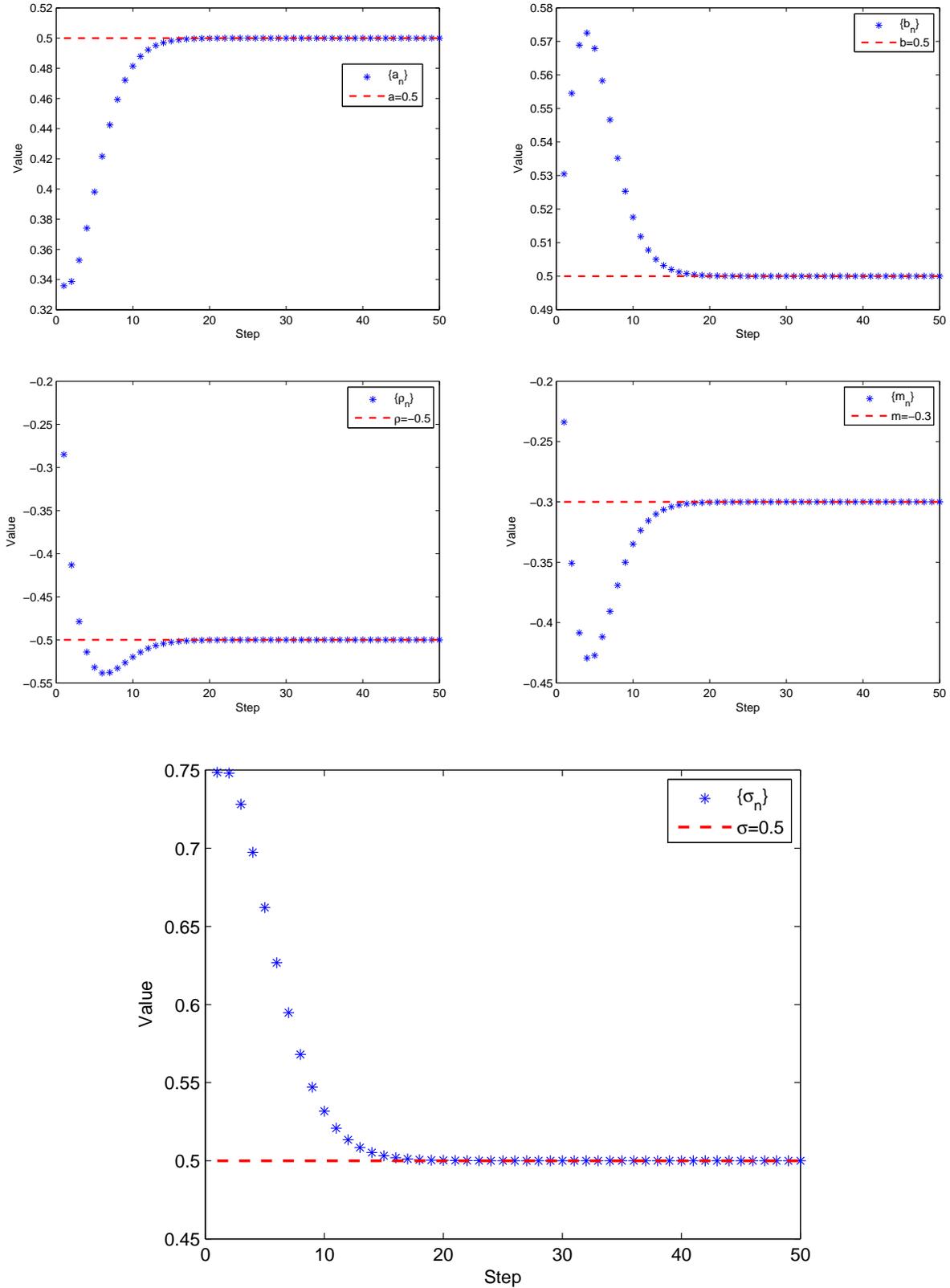

Based on the FPI-SVI algorithm, the estimations and root average squared error (RASE) of model (1.1) are presented in Figure 4.

Figure 4: FPI-SVI algorithm: the left picture shows the estimations results of parameters in model (1.1), and the right picture shows the RASE of estimations of $v$ in model (1.1).

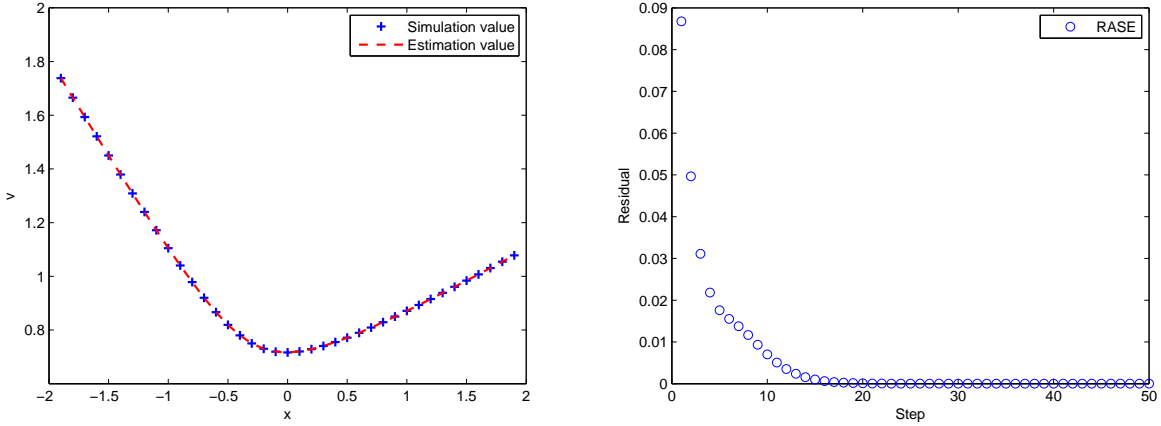

We conclude with the quasi-explicit SVI method of [15] as follows:

---
**Algorithm 2:** Quasi-explicit SVI method
---
**Input:** $(m_0, \sigma_0) = (x_{min}, v_{min})$, $\{x_i, v_i\}_{i=1}^{N}$

**Output:** Estimations of parameters $(a, b, \rho, m, \sigma)$

---

**Initialization**: $n = 0$, $M = 50$;
$$Y_0 = Y(m_0, \sigma_0);$$
$$\hat{\beta}_0 = \left[Y_0^\top Y_0\right]^{-1} Y_0^\top V.$$

---

**while** $n \leq M$ **do**
　$n = n + 1$;
　$(m_n, \sigma_n) = \min_{m,\sigma} \left(V - Y(m, \sigma)\hat{\beta}_{n-1}\right)^\top \left(V - Y(m, \sigma)\hat{\beta}_{n-1}\right)$;
　$Y_n = Y(m_n, \sigma_n)$;
　$\hat{\beta}_n = [Y_n^\top Y_n]^{-1} Y_n^\top V.$
**end**

---

We also perform 50 step iterations in the quasi-explicit SVI method and present the convergence property of sequences $\{a_n, b_n, \rho_n, m_n, \sigma_n\}_{n \geq 1}$ in Figure 5. Based on the quasi-explicit SVI method, the estimations and RASE of model (1.1) are recorded in Figure 6.

Figure 5: Quasi-explicit SVI method: values of sequences $\{a_n, b_n, \rho_n, m_n, \sigma_n\}_{n=1}^{50}$ along with the true values of parameters $(a, b, \rho, m, \sigma) = (0.5, 0.5, -0.5, -0.3, 0.5)$.

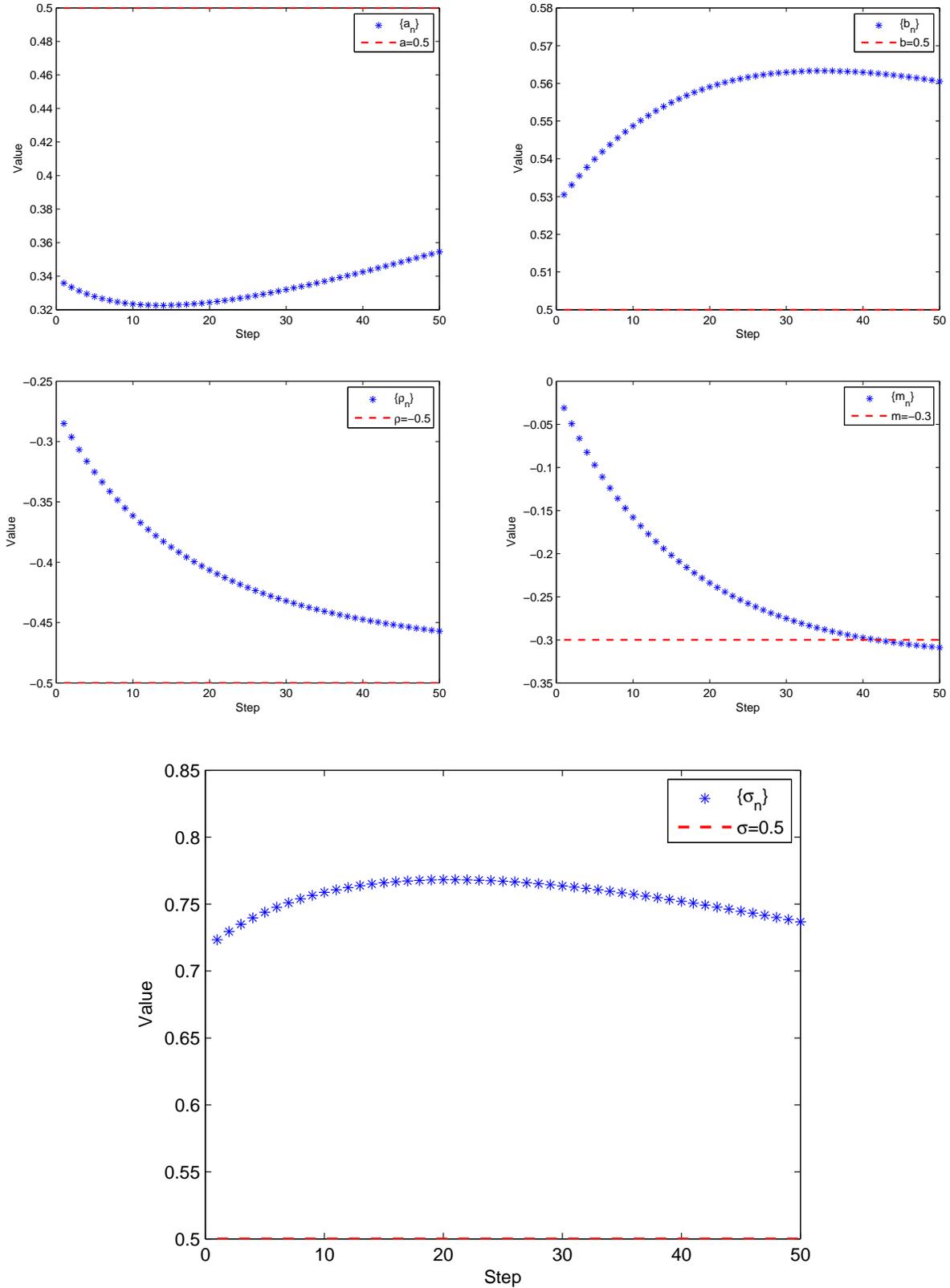

Figure 6: Quasi-explicit SVI method: the left picture shows the estimations results of parameters in model (1.1), and the right picture shows the RASE of estimations of *v* in model (1.1).

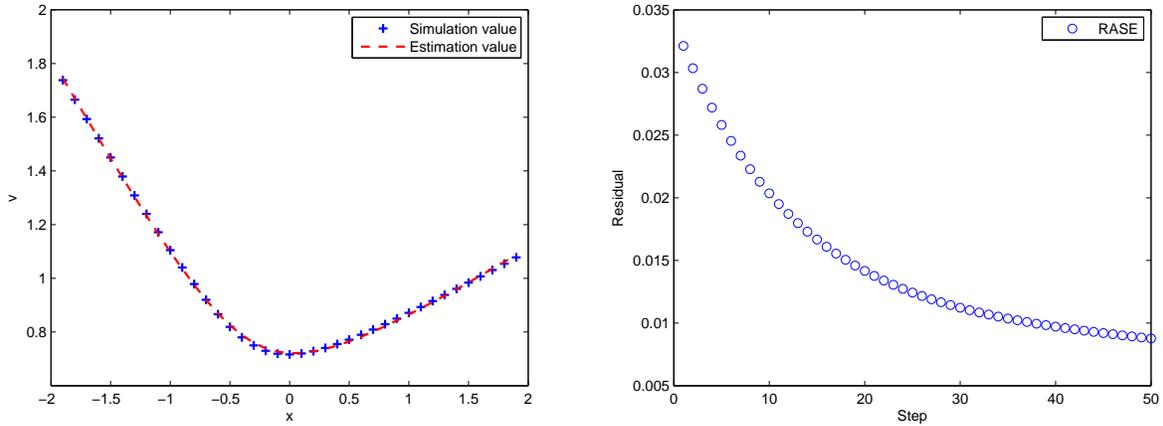

Based on the quasi-explicit SVI method, from Figure 6, the RASE of the estimations of *v* is 0.0088 (see Table 1) with 50 step iterations. To improve the accuracy of the quasi-explicit SVI method, we add the iterative steps of the quasi-explicit SVI method to 500 and estimate the parameters in 500-th step as $(a_n, b_n, \rho_n, m_n, \sigma_n)_{n=500} = (0.4992, 0.5004, -0.4998, -0.3002, 0.5015)$ which is almost identical to the true parameters $(0.5, 0.5, -0.5, -0.3, 0.5)$. For 500 iteration steps, the related RASE is 6.0895e-05, and the calculation time is 0.8482 s, which is based on a personal computer (PC). See Figure 7.

Figure 7: Quasi-explicit SVI method: the left picture shows the estimations results of parameters in model (1.1), and the right picture shows the RASE of estimations of *v* in model (1.1).

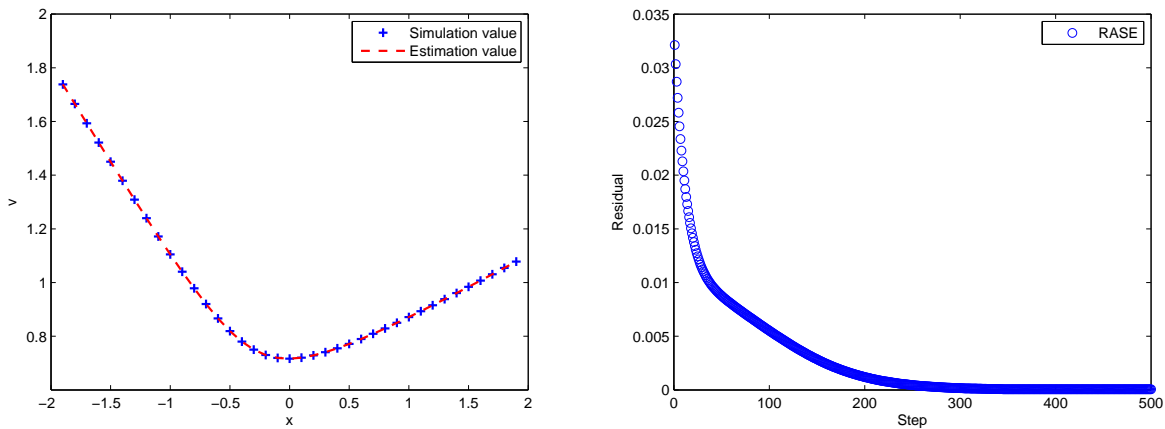

Table 1: Comparative of FPI-SVI algorithm and Quasi-explicit SVI (QE-SVI) with 50 steps.

|  | Method | $(a, b, \rho, m, \sigma)$ | RASE (RMSE) | Time |
|---|---|---|---|---|
| Case1 | True value | $(0.5000, 0.5000, -0.5000, -0.3000, 0.5000)$ | – – – | – – – |
|  | QE-SVI | $(0.3545, 0.5605, -0.4571, -0.3089, 0.7368)$ | 0.0088 (0.0157) | 0.0890 |
|  | FPI-SVI | $(0.5000, 0.5000, -0.5000, -0.3000, 0.5000)$ | 7.1450e-11 (2.0170e-10) | 0.0017 |
| Case2 | True value | $(0.0500, 0.6300, -0.5500, 0.0360, 0.2600)$ | – – – | – – – |
|  | QE-SVI | $(0.0403, 0.6344, -0.5433, 0.0396, 0.2756)$ | 0.0014 (0.0028) | 0.0856 |
|  | FPI-SVI | $(0.0500, 0.6300, -0.5500, 0.0360, 0.2600)$ | 4.8897e-16 (1.1102e-15) | 0.0016 |
| Case3 | True value | $(0.0500, 0.6300, 0.5500, 0.0360, 0.2600)$ | – – – | – – – |
|  | QE-SVI | $(0.0430, 0.6332, 0.5454, 0.0336, 0.2716)$ | 0.0011 (0.0021) | 0.0941 |
|  | FPI-SVI | $(0.0500, 0.6300, 0.5500, 0.0360, 0.2600)$ | 5.2685e-16 (1.3323e-15) | 0.0016 |
| Case4 | True value | $(0.1000, 0.0600, -0.7000, 0.2400, 0.0600)$ | – – – | – – – |
|  | QE-SVI | $(0.1000, 0.0600, -0.5999, 0.2401, 0.0601)$ | 2.7540e-06 (5.3243e-06) | 0.0821 |
|  | FPI-SVI | $(0.1000, 0.0600, -0.7000, 0.2400, 0.0600)$ | 5.0124e-16 (8.3267e-16) | 0.0015 |

# 4 Empirical analysis

In the following, we use financial market data to verify our FPI-SVI algorithm. Considering the implied variance smiles on the soybean meal option on Apr. 07, 2022, in China, there are eight contracts:1, m2205.DCE; 2, m2207.DCE; 3, m2209.DCE; 4, m2208.DCE; 5, m2211.DCE; 6, m2212.DCE; 7, m2301.DCE; and 8, m2303.DCE. We take m2205.DCE as an example to explain the details of the contracts: m2205.DCE denotes the maturity time of the underlying future contract in May. 2022, and the option's maturity time is the 5-th trading day of Apr. 2022. The other seven types of contracts provide similar explanations. Other Chinese market data implied variance smiles on copper options on Apr. 07, 2022; in China, there are three contracts:1, cu2205.SHFE; 2, cu2206.SHFE; and 3, cu2207.SHFE. We also consider four option contracts of American market with the underlying asset SPX (S&P500). Application of the FPI-SVI algorithm, we first conclude the estimation results and error RASE of contracts 1-4 of the soybean meal option in Figure 8, and the estimation results and error RASE of contracts 5-8 in Figure 9.

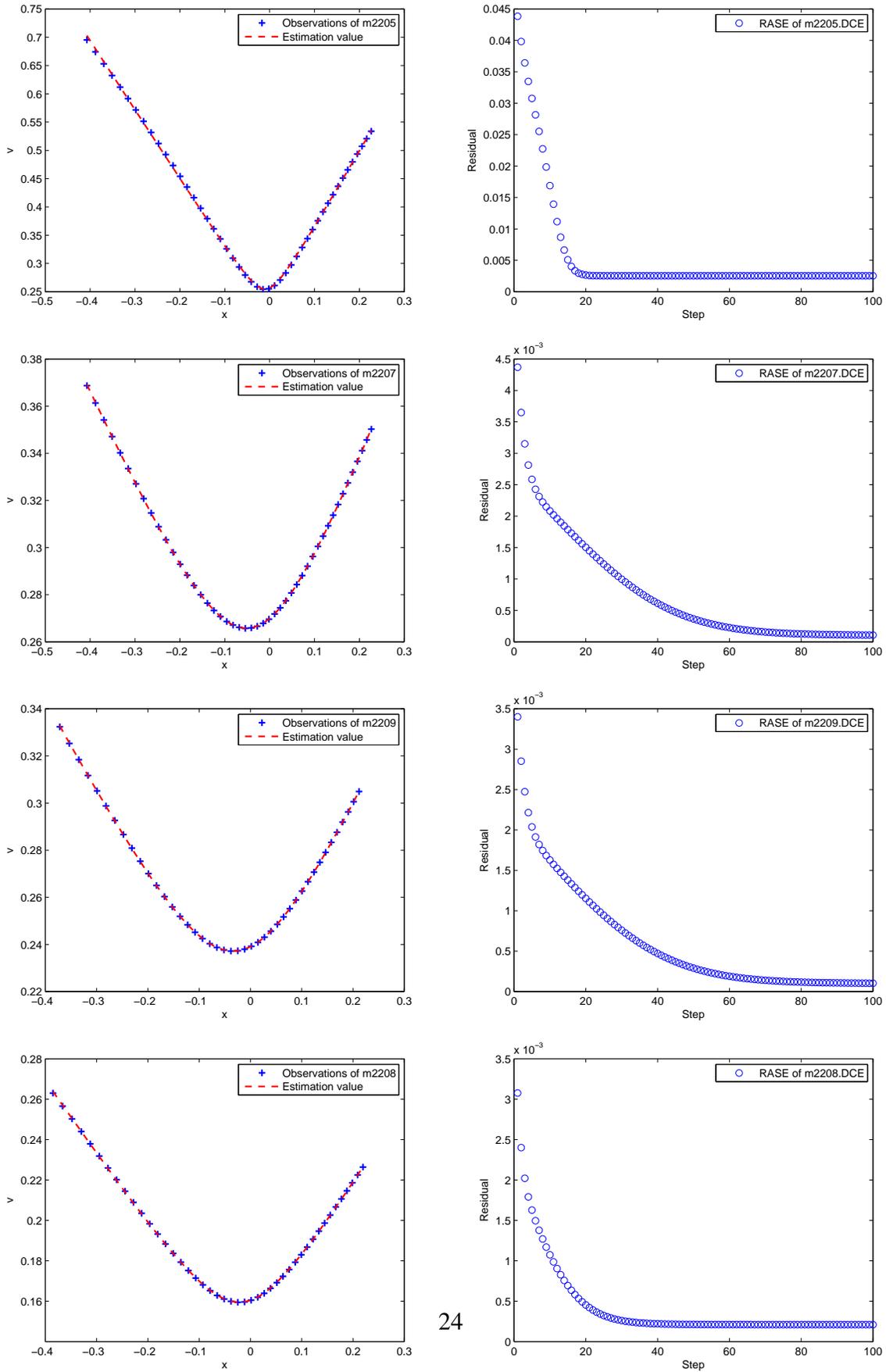

Figure 8: Estimations and RASE of 1-4 contacts of soybean meal option.

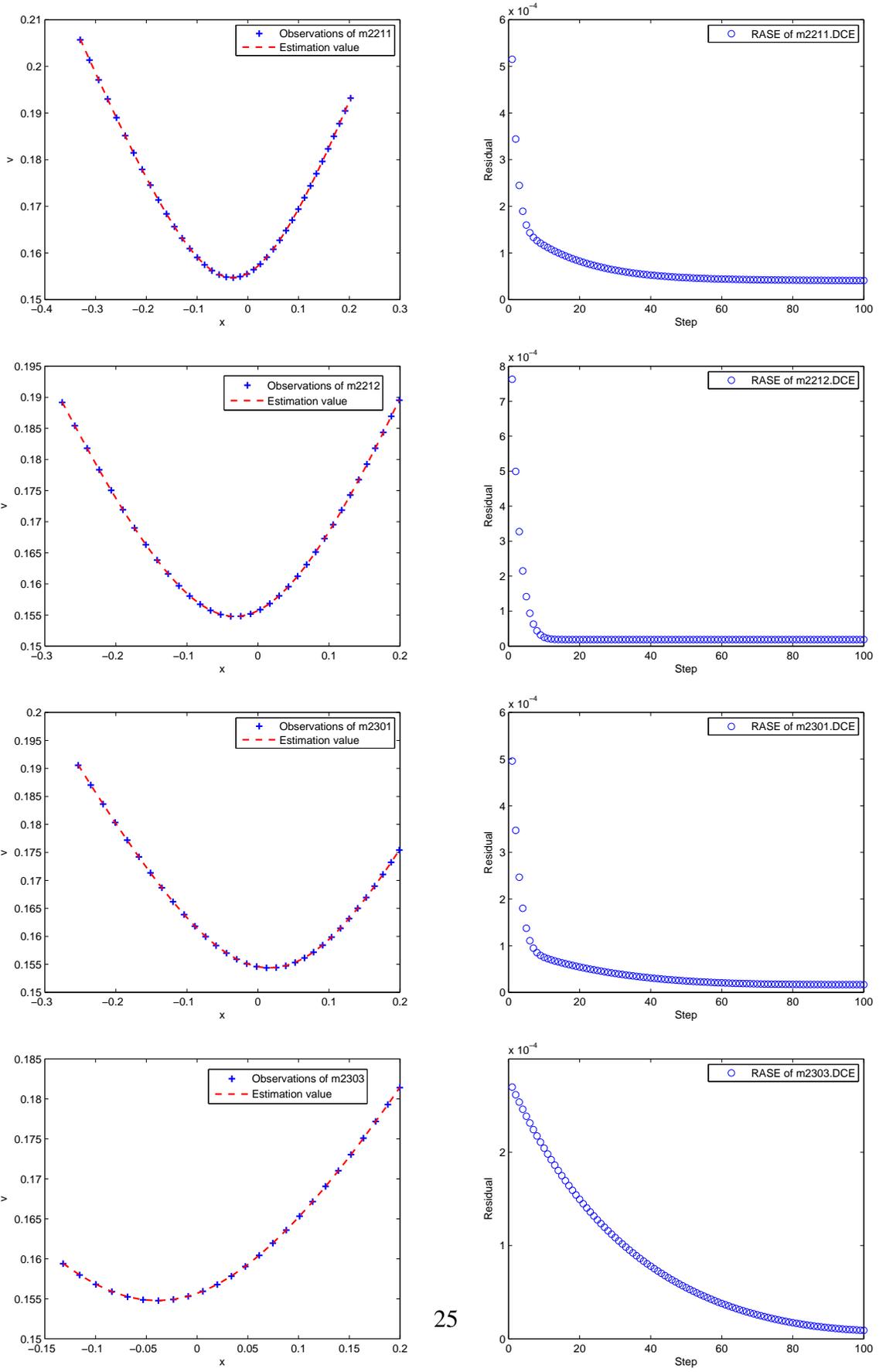

Figure 9: Estimations and RASE of 5-8 contacts of soybean meal option.

Figures 8 and 9 show that our FPI-SVI algorithm performs perfectly for the implied variance smile of the soybean meal option with 100 iteration steps. The RASE of the implied variance smile of the contract reduces to the error level of 1.0e-4, excluding contract 1. In the practice analysis, the error level of 1.0e-4 matches the accuracy of the requirements. If there are several points far away from the SVI model, we cannot obtain a sufficiently small RASE, as observed for contract 1. This phenomenon helps to locate a point that violates the no-arbitrage principle. However, a possible reason is that we have not found a better minimum point based on the observations of contract 1. We then use the quasi-explicit SVI method to verify these assertions. We consider 10000 steps of quasi-explicit SVI method for contract 1, and the RASE is 0.0021, which is almost identical to the RASE (0.0025) in the FPI-SVI algorithm with 100 iteration steps. Thus, it is reasonable to accept the estimations of the FPI-SVI algorithm.

Furthermore, from the values of RASE in Figures 8 and 9, FPI-SVI algorithm almost converges to the limit after the 50-th step. The quasi-explicit SVI method converges exceedingly slowly to the limit. We conclude the estimations of $(a, b, \rho, m, \sigma)$, RASE, and calculation time of the FPI-SVI algorithm and quasi-explicit SVI method for the eight contracts in Table 2 and Figure 10. In Table 2, we consider the FPI-SVI algorithm and quasi-explicit SVI method with 100 steps. We initially analyze the calculation time in Table 2. For the FPI-SVI algorithm, the calculation time of each contract is stable at approximately 0.003 s; for the quasi-explicit SVI method, the calculation time of each contract is stable at approximately 0.015 s. Thus, with the same calculation steps, the quasi-explicit SVI method requires a calculation time of 50 times that of the FPI-SVI algorithm.

We now analyze the RASE of the FPI-SVI algorithm and the quasi-explicit SVI method. Figure 10 shows the RASE of each contract under the FPI-SVI algorithm and quasi-explicit SVI methods. For all contracts 1-8, the RASE of the FPI-SVI algorithm is uniformly smaller than that of the quasi-explicit SVI method.

We now provide comments on the estimations of $(a, b, \rho, m, \sigma)$. For contracts 5-8, the estimations of $(a, b, \rho, m, \sigma)$ of the FPI-SVI algorithm and quasi-explicit SVI methods are almost identical. These results verify that the error level 1.0e-4 of RASE matches the accuracy of the requirements. For contracts 1-4, the RASE of the quasi-explicit SVI method is almost 1.0e-3. Thus, we should reject the estimations of $(a, b, \rho, m, \sigma)$ of the quasi-explicit SVI method and accept those of the FPI-SVI algorithm.

Table 2: Comparative of FPI-SVI algorithm and Quasi-explicit SVI method with 100 steps.

| Contract | Method | $(a, b, \rho, m, \sigma)$ | RASE (RMSE) | Time |
|---|---|---|---|---|
| (1) m2205 | QE-SVI | $(-0.2327, 2.3892, 0.2238, 0.0447, 0.2243)$ | 0.0191 (0.0358) | 0.1430 |
| | FPI-SVI | $(0.2117, 1.2957, 0.0528, -0.0084, 0.0329)$ | 0.0025 (0.0077) | 0.0036 |
| (2) m2207 | QE-SVI | $(0.0994, 0.6478, 0.1046, -0.0262, 0.2617)$ | 0.0017 (0.0042) | 0.1422 |
| | FPI-SVI | $(0.1966, 0.4719, 0.0828, -0.0393, 0.1469)$ | 1.0674e-04 (3.7943e-04) | 0.0034 |
| (3) m2209 | QE-SVI | $(0.1011, 0.5969, 0.1011, -0.0105, 0.2322)$ | 0.0013 (0.0033) | 0.1520 |
| | FPI-SVI | $(0.1755, 0.4475, 0.0626, -0.0252, 0.1380)$ | 1.0303e-04 (3.9441e-04) | 0.0035 |
| (4) m2208 | QE-SVI | $(0.0974, 0.4559, 0.0941, -0.0061, 0.1402)$ | 0.0011 (0.0022) | 0.1351 |
| | FPI-SVI | $(0.1212, 0.3937, 0.0502, -0.0168, 0.0974)$ | 2.0978e-04 (7.7273e-04) | 0.0045 |
| (5) m2211 | QE SVI | $(0.1100, 0.2932, 0.0513, -0.0228, 0.1530)$ | 1.1742e-04 (3.6059e-04) | 0.1325 |
| | FPI-SVI | $(0.1138, 0.2845, 0.0502, -0.0232, 0.1440)$ | 4.0938e-05 (1.5652e-04) | 0.0042 |
| (6) m2212 | QE-SVI | $(0.1127, 0.2710, 0.0413, -0.0269, 0.1555)$ | 5.2660e-05 (1.5188e-04) | 0.1661 |
| | FPI-SVI | $(0.1128, 0.2709, 0.0498, -0.0249, 0.1551)$ | 1.8714e-05 (5.4755e-05) | 0.0038 |
| (7) m2301 | QE-SVI | $(0.1176, 0.2390, 0.0299, 0.0202, 0.1533)$ | 7.0524e-05 (1.8351e-04) | 0.1557 |
| | FPI-SVI | $(0.1126, 0.2520, 0.0502, 0.0249, 0.1659)$ | 1.6512e-05 (6.2067e-05) | 0.0033 |
| (8) m2303 | QE-SVI | $(0.1229, 0.1991, 0.0618, -0.0285, 0.1597)$ | 9.6102e-05 (2.0094e-04) | 0.1761 |
| | FPI-SVI | $(0.1123, 0.2232, 0.0486, -0.0300, 0.1905)$ | 9.0554e-06 (1.6536e-05) | 0.0033 |

Figure 10: RASEs of FPI-SVI algorithm and Quasi-explicit SVI method.

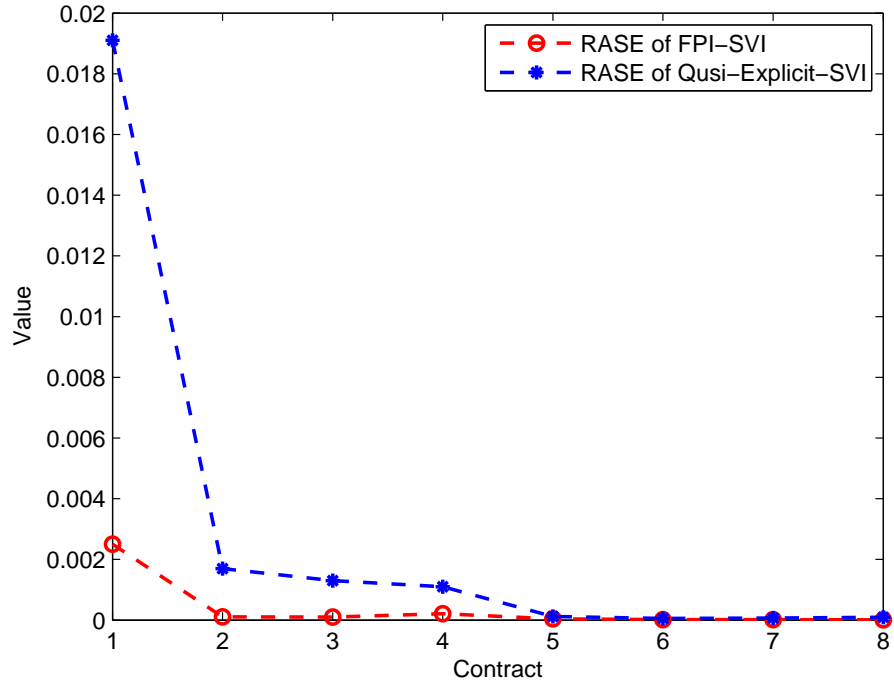

To further compare the FPI-SVI algorithm and the quasi-explicit SVI methods, we consider three contracts for the implied variance smiles of the copper option. The results in Table 3 lead to the same conclusion for the FPI-SVI algorithm and quasi-explicit SVI method, as presented in the implied variance smiles for the soybean meal option.

Table 3: Comparative of FPI-SVI algorithm and Quasi-explicit SVI method with 100 steps.

| Contract | Method | $(a, b, \rho, m, \sigma)$ | RASE (RMSE) | Time |
|---|---|---|---|---|
| (1) cu2205 | QE-SVI | $(-0.2172, 2.5535, 0.0854, -0.0020, 0.1482)$ | 0.0062 (0.0107) | 0.1307 |
| | FPI-SVI | $(0.1043, 1.3224, 0.0941, -0.0061, 0.0344)$ | 0.0035 (0.0064) | 0.0029 |
| (2) cu2206 | QE-SVI | $(-0.1036, 1.6181, -0.0601, -0.0391, 0.1680)$ | 9.5164e-04 (0.0018) | 0.1573 |
| | FPI-SVI | $(-0.1312, 1.6849, -0.0125, -0.0296, 0.1775)$ | 0.0013 (0.0024) | 0.0032 |
| (3) cu2207 | QE-SVI | $(-0.0006, 0.9977, -0.0174, -0.0375, 0.1766)$ | 8.0186e-04 (0.0018) | 0.1603 |
| | FPI-SVI | $(-0.0433, 1.1016, 0.0015, -0.0336, 0.1990)$ | 8.5023e-04 (0.0020) | 0.0031 |

Now, we consider four contracts with the underlying asset SPX (S&P500) on Feb. 15, 2019. The maturities are Mar. 15, 2019, June. 21, 2019, Sept. 20, 2019, and Dec. 20, 2019. At the beginning of this section, we have verified that the FPI-SVI algorithm has perfect performance, at least for the soybean meal option and the copper option in the Chinese financial market. In Table 4 and Figure 11, we show that the FPI-SVI algorithm is still useful for the stock index SPX option in the American financial market.

Table 4: Comparative of FPI-SVI algorithm and Quasi-explicit SVI method with 100 steps.

| Maturity | Method | $(a, b, \rho, m, \sigma)$ | RASE (RMSE) |
|---|---|---|---|
| Mar. 15 | QE-SVI | $(-0.0659, 1.6925, -0.1557, 0.0340, 0.1053)$ | 0.0019 (0.0075) |
| | FPI-SVI | $(0.0667, 0.9664, -0.0214, 0.0505, 0.0431)$ | 0.0015 (0.0057) |
| June. 21 | QE-SVI | $(0.0106, 0.9122, -0.0035, 0.1035, 0.1120)$ | 5.1605e-04 (0.0010) |
| | FPI-SVI | $(0.0822, 0.5568, -0.0473, 0.1005, 0.0536)$ | 1.9565e-04 (4.8999e-04) |
| Sept. 20 | QE-SVI | $(0.0402, 0.6469, -0.0103, 0.1358, 0.1138)$ | 2.8947e-04 (7.6152e-04) |
| | FPI-SVI | $(0.0713, 0.5057, -0.1123, 0.1262, 0.0847)$ | 1.9453e-04 (5.1154e-04) |
| Dec. 20 | QE-SVI | $(0.0565, 0.5143, -0.0046, 0.1593, 0.1165)$ | 1.9254e-04 (5.8892e-04) |
| | FPI-SVI | $(0.1212, 0.3937, 0.0502, -0.0168, 0.0974)$ | 5.3485e-04 (1.188e-03) |

Figure 11: FPI-SVI: Estimations and RASE of 4 contacts of SPX option.

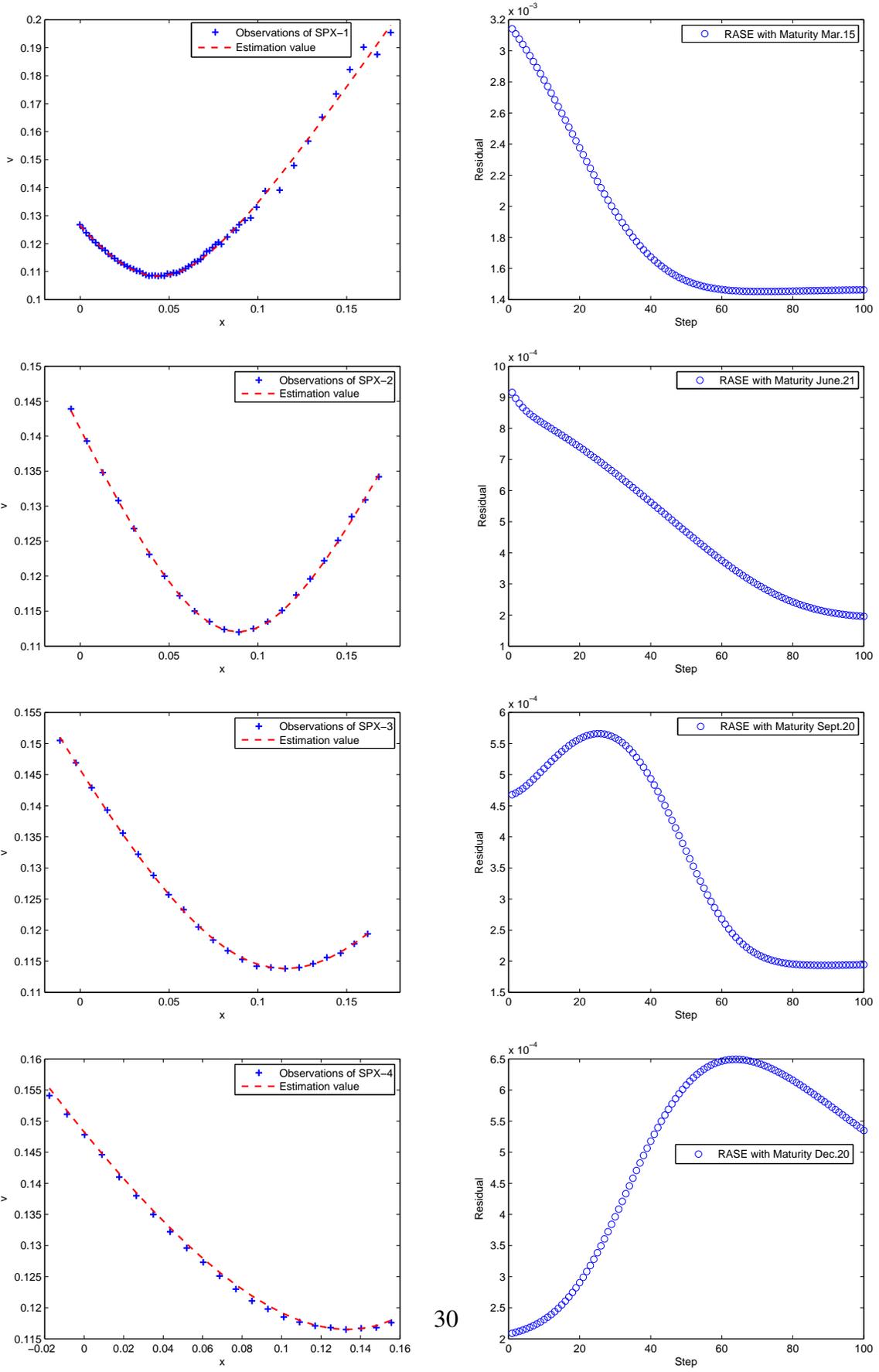

# 5  Conclusion

The SVI model, applied to describe implied volatility smiles, is widely used in financial markets because of its profound relationship with implied variance and excellent fit with observations. Presently, most optimizer algorithms of the SVI model strongly depend on the input starting point. We develop an explicit iterative algorithm that combines the minimum point in the SVI model and the least-squares optimization to establish a stable and efficient algorithm for the SVI model.

We establish some convergence results for the FPI-SVI algorithm under certain situations and demonstrate the excellent performance of this algorithm using simulation and market data. We also compared the FPI-SVI algorithm with the quasi-explicit SVI method regarding the accuracy of the estimations, convergence properties, and RASE (RMSE) of the parameters. The main result shows that using the same number of iterative steps, the quasi-explicit SVI method requires almost 50 times the calculation time of the FPI-SVI algorithm. Furthermore, the performance of the estimation of parameters and RASE (RMSE) of the FPI-SVI algorithm is better than that of the quasi-explicit SVI method, which implies the stability and efficiency of the FPI-SVI algorithm.

The FPI-SVI algorithm has several advantages in determining a better estimation of the parameters of the SVI model. We point out that the FPI-SVI algorithm cannot guarantee no arbitrage parameters of SVI model suggested by [12]. Developing no arbitrage SVI model is important for financial markets, we intend to study further in future works.

# References

[1] Y. Aït-Sahalia, C. Li and C. X. Li. Implied Stochastic Volatility Models. The Review of Financial Studies, 2021, 34(1), 394–450,

[2] Y. Aït-Sahalia, C. Li and C. X. Li. Closed-form implied volatility surfaces for stochastic volatility models with jumps. Journal of Econometrics, 2021, 222(1), 364–392.

[3] Z. Cui, J. Kirkby, D. Nguyen, S, Taylor. A closed-form model-free implied volatility formula through delta families. The Journal of Derivatives, 2021, 28(4), 111–127.

[4] M. Forde, A. Jacquier and A. Mijatović. Asymptotic formulae for implied volatility in the Heston model. Proc. R. Soc. A, 2010, 466, 3593–3620.

[5] F.Le Floch. Initial guesses for SVI calibration. SSRN-id2501898, 2014, 1–16.

[6] J. Gatheral. A parsimonious arbitrage-free implied volatility parameterization with application to the valuation of volatility derivatives, Presentation at Global Derivatives, 2004.

[7] J. Gatheral. The volatility surface: a practitioner's guide, 2006, 357, John Wiley & Sons.

[8] J. Gatheral, A. Jacquier. Convergence of Heston to SVI. Quantitative Finance, 2011, 11(8), 1129–1132.

[9] J. Gatheral, A. Jacquier. Arbitrage-free SVI volatility surfaces. Quantitative Finance, 2014, 14(1), 59–71.

[10] G. Guo, A. Jacquier, C. Martini, L. Neufcourt. Generalized arbitrage-free SVI volatility surfaces. SIAM J. Financial Mathematics, 2016, 7, 619–641.

[11] S. Hendriks, C. Martini. The extended SSVI volatility surface. J. Computational Finance, 2019, 22(5), 25–39.

[12] C. Martini, A. Mingone. No arbitrage SVI. SIAM Journal on Financial Mathematics, 2022, 13(1), 227–261.

[13] A. Mingone. No arbitrage global parametrization for the eSSVI volatility surface. arxiv: 2204.00312, 2022.

[14] R. W. Lee. The moment formula for implied volatility at extreme strikes. Mathematical Finance, 2004, 14(3), 469–480.

[15] Zeliade Systems. Quasi-Explicit Calibration of Gatheral's SVI model. 2009, Technical report, Zeliade White Paper.